\documentclass[aps,showpacs,pra,superscriptaddress,]{revtex4}
\usepackage{amsmath}
\usepackage{amsfonts}
\usepackage{amssymb}
\usepackage{bm}
\usepackage{graphicx}

\begin{document}

\title{Propagation of coupled quartic and dipole multi-solitons in optical
fibers medium with higher-order dispersions }
\author{Vladimir I. Kruglov}
\affiliation{Centre for Engineering Quantum Systems, School of Mathematics and Physics,
The University of Queensland, Brisbane, Queensland 4072, Australia}
\author{ Houria Triki}
\affiliation{Radiation Physics Laboratory, Department of Physics, Faculty of Sciences,
Badji Mokhtar University, P. O. Box 12, 23000 Annaba, Algeria}

\begin{abstract}
We present the discovery of two types of multiple-hump soliton modes in a highly dispersive optical fiber with a Kerr nonlinearity. We show that multi-hump optical solitons of quartic or dipole types are possible in the fiber system in the presence of higher-order dispersion. Such nonlinear wave
packets are very well described by an extended nonlinear Schr\"{o}dinger equation involving both cubic and quartic dispersion terms. It is found that the third- and fourth-order dispersion effects in the fiber material may lead to the coupling of quartic or dipole solitons into double-, triple-,
and multi-humped solitons. We provide the initial conditions for the formation of coupled multi-hump quartic and dipole solitons in the fiber. Numerical results illustrate that propagating multi-quartic and multi-dipole solitons in highly dispersive optical fibers councide with a high accuracy to our analytical multi-soliton solutions. It is important for applications that described multiple-hump soliton modes are stable to small noise perturbation that was confirmed by numerical simulations. These numerical results confirm that the newly found multi-soliton pulses can be potentially utilized for transmission in optical fibers medium with higher-order dispersions.
\end{abstract}

\pacs{05.45.Yv, 42.65.Tg}
\maketitle

\section{Introduction}

Nonlinear evolution of light pulses inside an optical fiber is affected by
group velocity dispersion and self-phase modulation. These two physical
effects can be exactly balanced in the transmitting medium leading to the
formation of optical solitons applicable to picosecond regime. We should
point out here that the self-phase modulation in silica glass fibers is the
nonlinear process arising from the lowest dominant third-order
susceptibility $\chi ^{(3)}$ \cite{Porsezian}. Dynamics of an envelope
soliton in a single-mode fiber is described by the one-dimensional cubic
nonlinear Schr\"{o}dinger equation (NLSE) which contains only basic effects
on waves, the linear group velocity dispersion and nonlinearily induced
self-phase modulation. Such equation supports the so-called
\textquotedblleft bright\textquotedblright\ and \textquotedblleft
dark\textquotedblright\ solitons in the anomalous dispersion (negative group
velocity dispersion) and normal dispersion (positive group velocity
dispersion) regimes, respectively. These optical solitons\ were
experimentally observed not only in monomode optical fibers \cite{E1}, but
also in many physical systems such as bulk optical materials \cite{E2} and
femtosecond lasers \cite{E3}. Nowadays, optical solitons are attracting
great interest with a view to their extensive applications in long-distance
communications, pulse shaping in laser sources, and optical switching
devices \cite{Akhmediev}. An interesting property of these localized wave
packets is their remarkable robustness displayed in their propagation and
interaction \cite{Stegeman}.

But with injecting short pulses (to nearly $50$ fs) in optical waveguiding
media, the effect of third-order dispersion becomes significant and should
be incorporated in the underlying equation \cite{Palacios1}. In addition, as
the pulses become extremely short (below $10$ fs), fourth-order dispersion
should also be taken into consideration \cite{Palacios2,Palacios3}. To
describe the propagation of such high-power femtosecond light pulses inside
the optical system, more generalized NLSE models containing various
contributions of higher-order dispersive and nonlinear terms have been
developed, giving rise to subpicosecond solitons.

A very special class of optical solitons is the so-called \textit{quartic
solitons} which are formed by the interplay between self-phase modulation
and anomalous second- and fourth-order dispersions. Studies of such
shape-maintaining nonlinear wave packets constitute a very active field \cite%
{H1,H2,H3,H4,H5,H6,P}, with a potential for practical applications to
optical communications and photonics. The current availability of silicon
photonics has opened a new way to generate waveguide structures exhibiting a
wide range of dispersion profiles which allow possible quartic soliton
creation \cite{S1,S2,S3,S4,S6}. In this regard, impressive results obtained
recently indicate the possibility of observing localized quartic solitons in
specially designed silicon-based slot waveguides \cite{R1}. Additionally, a
detailed analysis has demonstrated the existence of a stable quartic soliton
envelope taking the form of $\mathrm{sech}^{2}$ in optical fibers exhibiting
all orders of dispersion up to the order four \cite{K,Kr}. It is interesting
to note that quartic solitons can be generated not only in presence of
higher-order dispersions but also under the influence of self-steepening
process \cite{K1}.

Very recently, we reported the formation of stable dipole solitons in an
optical fiber medium exhibiting all orders of dispersion up to the order
four \cite{T}. The obtained results showed that such dipole-mode solitons
characteristically exist due to a balance among the effects of second-,
third-, and fourth-order dispersions and self-phase modulation. Notice that
a dipole soliton is a localized structure possessing two symmetrical humps
with a zero intensity value in the middle of the pulse \cite{Chou}. It is
also interesting to note that dipole-mode solitons have been recently
observed in a three-level cascade atomic system \cite{Yanpeng}. It seems
natural to ask whether there is a possibility of formation of stable
multi-hump solitons in an optical Kerr medium exhibiting second-, third-,
and fourth-order dispersions? Identifying this kind of solitons in a
waveguide system is a challenging problem that requires extensive numerical
simulations. Such multi-hump solitons are composed of several coupled
solitons possessing the appropriate symmetry. It is relevant to mention that
the first experimental observation of a multihumped multimode bright soliton
in a dispersive nonlinear medium was presented in \cite{Matthew}.

In this paper, we present the first demonstration of the existence of multi-hump 
\textit{quartic} and \textit{dipole} solitons in a highly dispersive optical
fiber system. We show analytically and numerically that considering the
combined influence of third- and fourth-order dispersion effects may pave
the way to generate multiple soliton pulses in the form of coupled two, three and four solitons and also multi-quartic or multi-dipole N-solitons. Importantly, such wave packets are
very well described by the so-called extended nonlinear Schr\"{o}dinger
equation (NLSE) incorporating high order of dispersion and Kerr nonlinearity. The newly found multiple-hump solitons may further expand the applicalibity of quartic and dipole solitons for transmission in optical fiber.

The paper is organized in the following way. In Sect. II, the extended NLSE
model describing pulse propagation in a highly dispersive optical fiber and
its general coupled multi-soliton solutions are presented. In Sect. III, multi-quartic and multi-dipole N-soliton solutions are found. In Sect.
IV, the numerical simulations are performed to study the formation of
multi-hump solitons involving well separated two, three and four quartic or
dipole solitons in the system. In Sect. V, the stability properties of the
solutions is discussed. Finally, the conclusions are presented in Sec. VI.

\section{Model and multi-solitons in highly dispersive optical fiber}

Ultrashort light pulse transmission through a highly dispersive optical
fiber with Kerr-type nonlinearity is governed by the extended NLSE model 
\cite{P,K,Kr,T}: 
\begin{equation}
i\frac{\partial \psi }{\partial z}=\alpha \frac{\partial ^{2}\psi }{\partial
\tau ^{2}}+i\sigma \frac{\partial ^{3}\psi }{\partial \tau ^{3}}-\epsilon 
\frac{\partial ^{4}\psi }{\partial \tau ^{4}}-\gamma \left\vert \psi
\right\vert ^{2}\psi ,  \label{1}
\end{equation}%
where $\psi (z,\tau )$ is the complex field envelope, $z$ represents the
distance along the direction of propagation, and $\tau =t-\beta _{1}z$ is
the retarded time in the frame moving with the group velocity of wave
packets. Also $\alpha =\beta _{2}/2$, $\sigma =\beta _{3}/6$, and $\epsilon
=\beta _{4}/24$, with $\beta _{k}=(d^{k}\beta /d\omega ^{k})_{\omega =\omega
_{0}}$ denotes the k-order dispersion of the optical fiber with $\beta
(\omega )$ is the propagation constant depending on the optical frequency.
The parameter $\gamma $ represents the cubic nonlinearity coefficient.

In the absence of third- and fourth-order dispersion effects ($\sigma
=\epsilon =0$), the propagation equation (\ref{1}) reduces to the standard
NLSE, which is completely integrable by the inverse scattering transform 
\cite{Kodama}. For $\alpha =\sigma =0$, the model equation (\ref{1}) becomes
the biharmonic NLSE governing the propagation dynamics of pure-quartic
solitons in a silicon photonic crystal waveguide \cite{Blanco}. To date, no
multi-hump soliton solution of the quartic and dipole kind has been provided
for this physically important model. The characteristics of these types solitons will be the main objectives of
the present study.

To start with, we present the first analysis of the existence of
multi-soliton solutions of Eq. (\ref{1}), which includes higher-order
dispersion terms. So, we consider the case when $U$ and $V$ are the
localized solutions of Eq. (\ref{1}) and the sum $U+V$ is also the solution
of NSLE. The substitution of these functions to Eq. (\ref{1}) leads to three
NSLE with $\psi =U$, $\psi =V$ and $\psi =U+V$. Using some transformations
with these three NLSE one can find the condition that $U+V$ is the solution
of Eq. (\ref{1}) assuming the localized waves $U$ and $V$ are the solution
of this NLSE as well. This condition is given by the following algebraic
equation, 
\begin{equation}
U^{2}V^{\star }+2VUU^{\star }+2UVV^{\star }+V^{2}U^{\star }=0,  \label{2}
\end{equation}%
where the star means complex conjugation. Let the traveling localized waves $%
U$ and $V$ have the form as $U=f(\xi -\xi _{1})e^{i\phi (z,\tau )}$ and $%
V=g(\xi -\xi _{2})e^{i\theta (z,\tau )}$ where $f(\xi -\xi _{1})$ and $g(\xi
-\xi _{2})$ are real function of the variable $\xi =\tau -qz$. Here the
parameter $q=1/v$ is inverse velocity given by $q=\sigma (\sigma
^{2}-4\alpha \epsilon )/8\epsilon ^{2}$. The condition in Eq. (\ref{2}) is
satisfied when the following relation $f(\xi -\xi _{1})g(\xi -\xi _{2})=0$
holds. However, for interacting solitons forming multi-soliton pulses the
above condition is satisfied approximately only. Nevertheless the above
consideration leads to approach for coupled multi-soliton pulses based on
summation of the soliton solutions of NLSE (\ref{1}) with appropriate
constant phases.

The soliton solution of Eq. (\ref{1}) can be written as $\psi(x,\tau)=F(\xi-%
\xi_{0})\exp(i\Phi(z,\tau))$ and the coupled multi-soliton waves
(N-solitons) have the approximate form, 
\begin{equation}
\psi_{N}(z,\tau) =\sum_{n=1}^{N}F(\xi-\xi_{n})\exp[i(\Phi(z,\tau)+\Phi_{n})],
\label{3}
\end{equation}%
where $\xi_{n}=\xi_{0}+(n-1)a$ and $\Phi_{n}$ are the constant phases. The
parameter $a$ can be found using appropriate variational procedure. The
coupled N-soliton solution given in Eq. (\ref{3}) has a good precision when
the following condition $\max F^{2}(\xi)\gg w_{nm}$ (with $n\neq m$) is
satisfied. The numbers $w_{nm}$ in this inequality are given by 
\begin{equation}
w_{nm}=\max |F(\xi-\xi_{n})F(\xi-\xi_{m})|.  \label{4}
\end{equation}%
Note that the condition $\max F^{2}(\xi)\gg w_{nm}$, without loss of generality, can also be
written as 
\begin{equation}
\max F^{2}(\xi)\gg w_{12}.  \label{5}
\end{equation}%
It is important that this condition is satisfied for multi-solitons propagating in optical
fibers medium with higher-order dispersions described by extended NLSE model (\ref{1}).
The wave function of coupled multi-soliton solution can be written as $%
\psi_{N}(z,\tau)=U_{N}(z,\tau)\exp(i\Phi(z,\tau))$ where $U_{N}(z,\tau)$ is
a real function. Moreover, this function is changing the sign at some point
between two neighbouring solitons of the coupled multi-soliton. Hence, the
wave function $\psi_{N}(z,\tau)$ of coupled N-soliton solution is equal to
zero at such points. Thus, the coupled N-soliton solution has $N-1$ ``zero
points". We emphasize that these ``zero points" are connected with stability
of the coupled N-soliton solutions of Eq. (\ref{1}). We show below that the
existence of these ``zero points" allow us to find the constant phases $%
\Phi_{n}$ for coupled multi-quartic and multi-dipole N-solitons.

\section{Coupled multi-quartic and multi-dipole solitons}

The approximate wave function of multi-quartic N-soliton is given in Eq. (\ref{3}) where the wave function of quartic soliton \cite{K,Kr,T} is 
\begin{equation}
\psi(x,\tau)=A_{0}\mathrm{sech}^{2}(w\xi)\exp[i(\kappa z-\delta\tau+\theta)],
\label{6}
\end{equation}%
where $\xi=\tau-qz$, and the velocity of quartic solitons in the moving frame is $v=1/q$. The amplitude $A_{0}$ and inverse width $w$ of quartic soliton are 
\begin{equation}
A_{0}=\pm\sqrt{\frac{-3}{10\gamma\epsilon}} \left(\frac{3\sigma^{2}-8\alpha%
\epsilon}{8\epsilon}\right),~~~~w=\frac{1}{4}\sqrt{\frac{8\alpha\epsilon-3%
\sigma^{2}}{10\epsilon^{2}}}.  \label{7}
\end{equation}%
The multi-quartic N-soliton is the coupling of quartic solitons $\psi(x,\tau) $ into N-soliton with constant phases $\Phi_{n}=(n-1)\pi$. Eq. (\ref{3}) yields in this case the multi-quartic N-soliton as 
\begin{equation}
\psi_{N}(z,\tau) =\sum_{n=1}^{N}(-1)^{n-1}A_{0}\mathrm{sech}%
^{2}(w(\xi-\xi_{n}))\exp[i(\kappa z-\delta\tau+\theta)],  \label{8}
\end{equation}%
where $\xi_{n}=\xi_{0}+(n-1)a$. The parameter $a=a_{0}w^{-1}$ is period of intensity ($I=|\psi_{N}|^{2}$) for quartic N-soliton where $a_{0}$ is the dimensionless constant which can be found theoretically or numerically.
We have evaluated the dimensionless constant $a_{0}$ for multi-quartic soliton numerically as $a_{0}=3$.
Note that the parameters $a$ and $a_{0}$ are defined by interaction of quartic solitons which constitute the coupled multi-quartic soliton given in Eq. (\ref{8}). The
found constant phases $\Phi_{n}=(n-1)\pi$ follow from the condition that real function $U_{N}(z,\tau)=\psi_{N}(z,\tau)\exp[-i(\kappa
z-\delta\tau+\theta)]$ in Eq. (\ref{8}) has $N-1$ ``zero points" located between neighbouring solitons of the coupled multi-quartic N-soliton. The frequency shift in the phase $\Phi(z,\tau)=\kappa z-\delta\tau+\theta$ of
multi-quartic N-soliton is $\delta=-\sigma/4\epsilon$ and the wave number $\kappa$ is given as 
\begin{equation}
\kappa=-\frac{4}{25\epsilon^{3}}\left(\frac{3\sigma^{2}}{8}
-\alpha\epsilon\right)^{2}-\frac{\sigma^{2}}{16\epsilon^{3}} \left(\frac{3\sigma^{2}}{16}-\alpha\epsilon\right).  \label{9}
\end{equation}
The variable $\xi=\tau-qz$ of quartic N-soliton solutions depends on velocity $v=1/q$ which is 
\begin{equation}
v=\frac{8\epsilon^{2}}{\sigma(\sigma^{2}-4\alpha\epsilon)}.  \label{10}
\end{equation}
We note that the quartic soliton given by Eq. (\ref{6}) has the same velocity $v$ as multi-quartic soliton in Eq. (\ref{8}). It follows from Eq. (\ref{7}) that multi-quartic N-solitons exist when the following conditions are satisfied: 
\begin{equation}
8\alpha\epsilon-3\sigma^{2}>0,~~~~\gamma\epsilon<0.  \label{11}
\end{equation}

The approximate wave function of multi-dipole N-soliton is presented in Eq. (\ref{3}) where the wave function of dipole soliton \cite{T} is 
\begin{equation}
\psi(x,\tau)=A_{0}\mathrm{sech}(w\xi)\mathrm{th}(w\xi) \exp[i(\kappa
z-\delta\tau+\theta)],  \label{12}
\end{equation}%
where $\xi=\tau-qz$, and the velocity of dipole solitons in the moving frame is $v=1/q$. The amplitude $A_{0}$ and inverse width $w$ of dipole soliton are 
\begin{equation}
A_{0}=\pm\sqrt{\frac{6}{5\gamma\epsilon}} \left(\frac{3\sigma^{2}-8\alpha%
\epsilon}{8\epsilon}\right),~~~~w=\frac{1}{4}\sqrt{\frac{3\sigma^{2}-8\alpha%
\epsilon}{5\epsilon^{2}}}.  \label{13}
\end{equation}%
The multi-dipole N-soliton is the coupling of dipole solitons $\psi(x,\tau)$ into N-soliton with constant phases $\Phi_{n}=0$. Eq. (\ref{3}) yields in this case the multi-dipole N-soliton as 
\begin{equation}
\psi_{N}(z,\tau) =\sum_{n=1}^{N}A_{0}\mathrm{sech}(w(\xi-\xi_{n}))\mathrm{th}%
(w(\xi-\xi_{n}))\exp[i(\kappa z-\delta\tau+\theta)],  \label{14}
\end{equation}%
where $\xi_{n}=\xi_{0}+(n-1)a$. The parameter $a=a_{0}w^{-1}$ is period of intensity for dipole N-soliton where $a_{0}$ is the dimensionless constant. We have evaluated the dimensionless constant $a_{0}$ for multi-dipole soliton numerically as $a_{0}=3.5$. The parameters $a$ and $a_{0}$ are defined by interaction of solitons which constitute the coupled multi-dipole soliton given in Eq. (\ref{14}). The constant phases $\Phi_{n}=0$ in this solution follow from the condition that real function $U_{N}(z,\tau)=\psi_{N}(z,\tau)\exp[-i(\kappa
z-\delta\tau+\theta)]$ in Eq. (\ref{14}) has $N-1$ ``zero points" located between neighbouring solitons of the coupled multi-dipole N-soliton. The frequency shift in the phase $\Phi(z,\tau)=\kappa z-\delta\tau+\theta$ of
multi-dipole N-soliton is $\delta=-\sigma/4\epsilon$ and the wave number $\kappa$ is given as 
\begin{equation}
\kappa=\frac{11}{100\epsilon^{3}}\left(\frac{3\sigma^{2}}{8}
-\alpha\epsilon\right)^{2}-\frac{\sigma^{2}}{16\epsilon^{3}} \left(\frac{3\sigma^{2}}{16}-\alpha\epsilon\right).  \label{15}
\end{equation}
The variable $\xi=\tau-qz$ in these dipole N-soliton solutions depends on velocity $v=1/q$ which is 
\begin{equation}
v=\frac{8\epsilon^{2}}{\sigma(\sigma^{2}-4\alpha\epsilon)}.  \label{16}
\end{equation}
We note that the dipole soliton given by Eq. (\ref{12}) has the same velocity $v$ as multi-dipole soliton in Eq. (\ref{14}). It follows from Eq. (\ref{13}) that multi-dipole N-solitons exist when the following conditions are satisfied: 
\begin{equation}
8\alpha\epsilon-3\sigma^{2}<0,~~~~\gamma\epsilon>0.  \label{17}
\end{equation}

We emphasize that the velocity $v$ of quartic solitons and multi-quartic solitons as well as dipole solitons and multi-dipole solitons given by Eqs. (\ref{10}) and (\ref{16}) coincide only in the form because the dispersion parameters as $\alpha, \sigma$ and $\epsilon$ in these two cases belong to
different domain presented by Eqs. (\ref{11}) and (\ref{17}) respectively. Moreover, the velocities in Eqs. (\ref{10}) and (\ref{16}) are fixed by dispersion parameters $\alpha, \sigma$ and $\epsilon$ which is important for stability of coupled multi-quartic and multi-dipole solitons. Using the numerical stability analysis we show in Sec.V that the multi-quartic and
multi-dipole solitons are stable. We also note that the multi-quartic and multi-dipole solitons given in Eqs. (\ref{8}) and (\ref{14}) coincide with a high accuracy with numerical multi-solitons presented in the Sec.IV. This result is connected with the condition given in Eq. (\ref{5}) which is satisfied for the novel multi-quartic and multi-dipole solitons in Eqs. (\ref{8}) and (\ref{14}) respectively.

\section{Numerical multi-hump solutions for extended NLSE}

We now find the numerical multi-hump solutions to the extended NLSE model (%
\ref{1}) by using the split-step Fourier method \cite{Agrawal}. Here, we
show that the cubic and quartic dispersions when acting in combination lead
to the coupling of quartic and dipole solitons into multiple-modes. We also
show that the shape of the newly found solutions depends crucially on the
choice of the initial condition. 
\begin{figure}[h]
\includegraphics[width=1\textwidth]{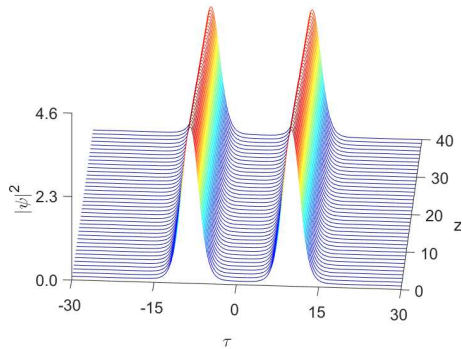}
\caption{Propagation of the double-hump quartic soliton solution of Eq. (1)
with parameters $\protect\alpha =-2.0375,$\ $\protect\gamma =2.6,$ $\protect%
\sigma =0.1,$ and $\protect\epsilon =-0.1$.}
\label{FIG.1.}
\end{figure}
\begin{figure}[h]
\includegraphics[width=1\textwidth]{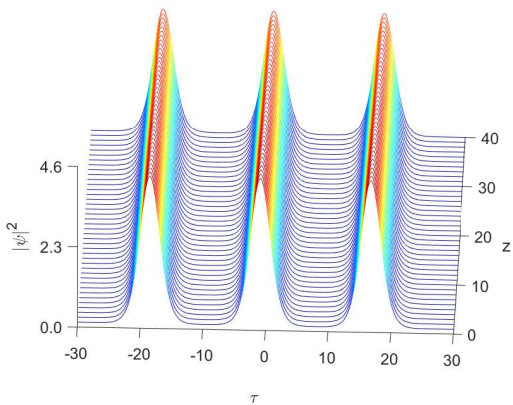}
\caption{Propagation of the triple-hump quartic soliton solution of Eq. (1).
The parameters are the same as in Fig. 1.}
\label{FIG.2.}
\end{figure}
First, we analyze the existence of multiple-hump quartic soliton solutions
of Eq. (\ref{1}) and demonstrate that waveforms of double-, triple-, and
quadruple-hump form can readily be generated in the fiber system. To perform
a numerical study of multi-hump quartic solitons, we numerically integrated
the full underlying equation (\ref{1}) using the analytic soliton solution (\ref{8}) with $N=2$ as an initial condition. The propagation of the
double-hump quartic solitons calculated within the framework of Eq. (\ref{1}%
) is shown in Fig. 1 for the parameter values: $\alpha =-2.0375,$\ $\gamma
=2.6,$ $\sigma =0.1,$ $\epsilon =-0.1,$ and $\xi _{0}=0.$\ It should be
noted that here, we have evaluated the dimensionless constant $a_{0}$
numerically as $a_{0}=3.$ From this figure, we see that the two humps are
well separated and both of them have the same shape (width and maximum
intensity). We also see that the wave profile remains unchanged after
propagating a distance of forty normalized lengths. Assuming initial conditions in the form of the analytic soliton solution (\ref{8}) containing
three ($N=3$) or four ($N=4$) superimposed single-hump \textrm{sech}$^{2}$
solitons, we observe that both tripole and quadrupole quartic solitons are
obtained, as shown in Figs. 2 and 3 respectively. 
\begin{figure}[h]
\includegraphics[width=1\textwidth]{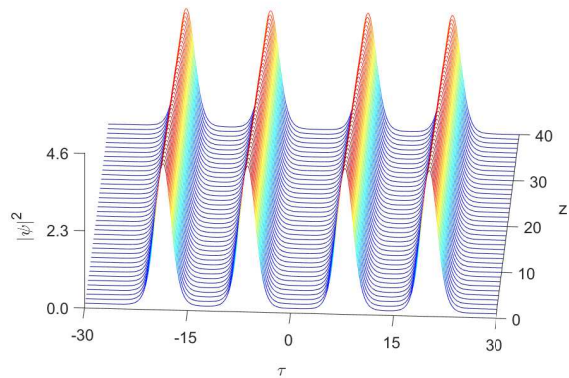}
\caption{Propagation of the quadruple-hump quartic soliton solution of Eq. (1). The parameters are the same as in Fig. 1.}
\label{FIG.3.}
\end{figure}
\begin{figure}[h]
\includegraphics[width=1\textwidth]{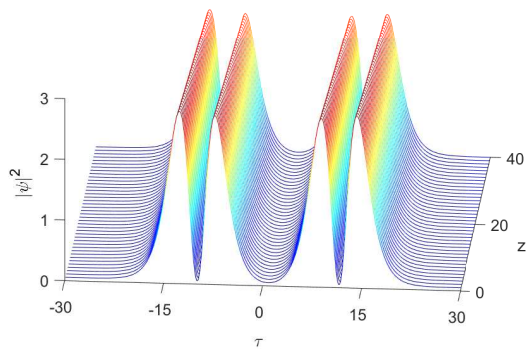}
\caption{Propagation of the double-hump dipole soliton solution of Eq. (1)
with parameters $\protect\alpha =-1,$\ $\protect\gamma =-2.6,$ $\protect%
\sigma =1.53,$ and $\protect\epsilon =-0.25$.}
\label{FIG.4.}
\end{figure}
This also indicates that these numerical findings agree excellently with our
analytic solution (\ref{8}). This physically important result implies that
such multi-quartic solitons can be observed experimentally as long as the
model (1) applies, thus implying that the solutions obtained here can be
utilized for transmission. 
\begin{figure}[h]
\includegraphics[width=1\textwidth]{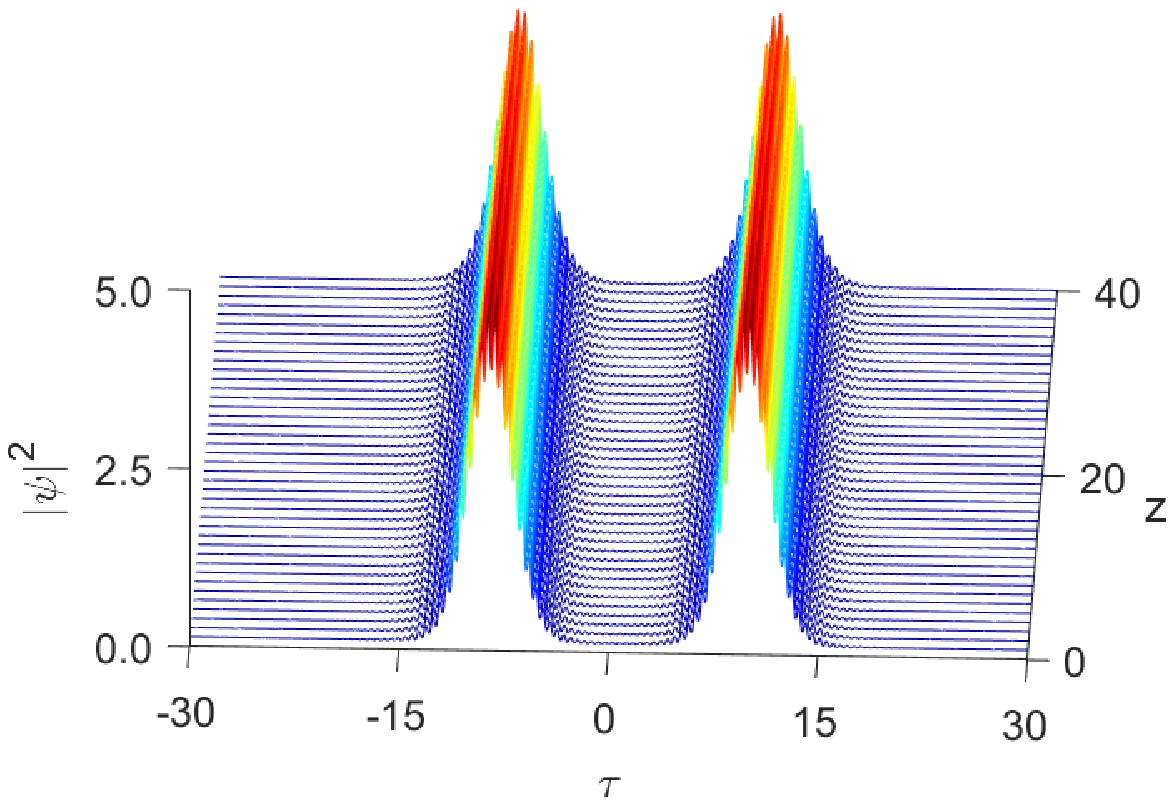}
\caption{The numerical evolution of double-hump quartic solitons under the
perturbation of white noise whose maximal value is $0.1$. The parameters are
the same as those used in Fig. 1.}
\label{FIG.5.}
\end{figure}
\begin{figure}[h]
\includegraphics[width=1\textwidth]{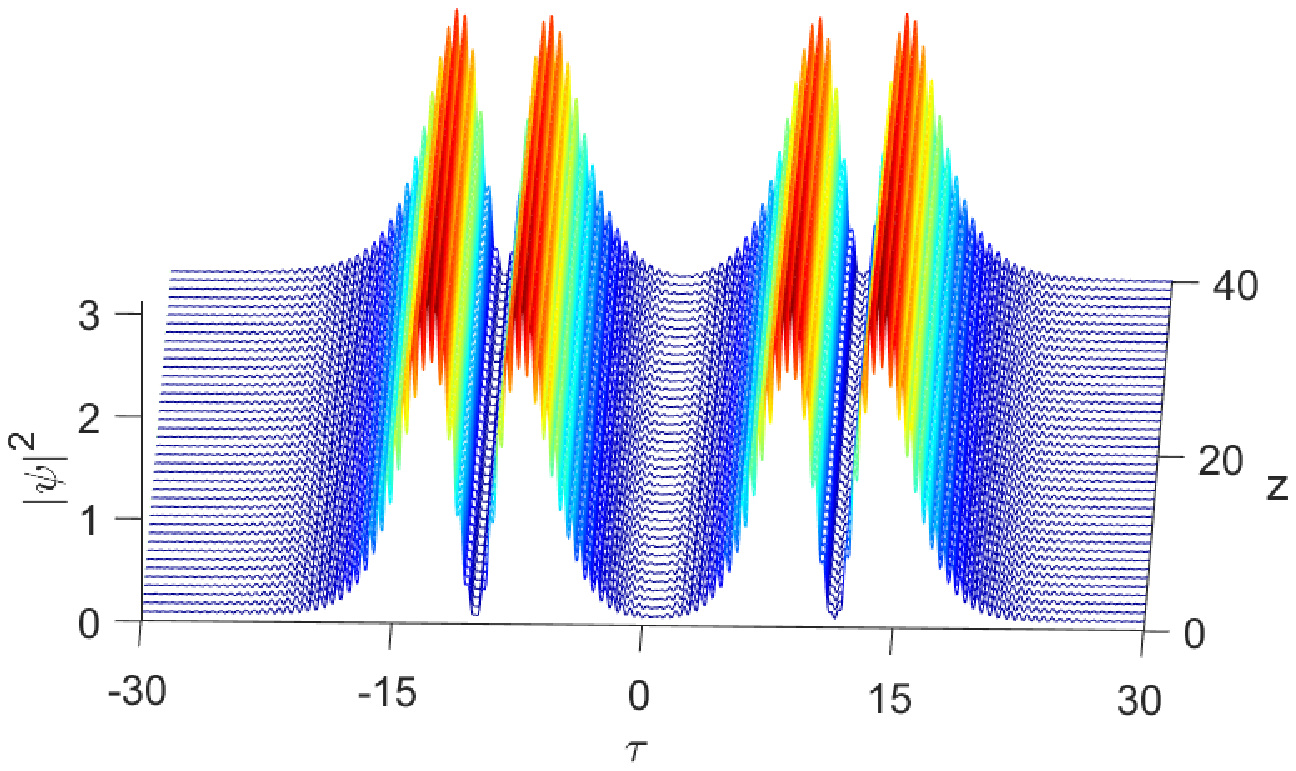}
\caption{The numerical evolution of double dipole solitons under the
perturbation of white noise whose maximal value is $0.1$. The parameters are
the same as those used in Fig. 4.}
\label{FIG.6.}
\end{figure}

Now we focus on the formation of the multiple-dipole soliton families in the
fiber medium given analytically by Eq. (\ref{14}). To obtain the numerical multi-hump solutions, we solved Eq. (\ref{1}) by means of the split-step Fourier method for an input condition
taking the form of the analytic soliton solution (\ref{14}) with $N=2$. Here
we have taken the following parameter values: $\alpha =-1,$\ $\gamma =-2.6,$ 
$\sigma =1.53,$ $\epsilon =-0.25,$ and $\xi _{0}=0.$ We have also calculated
the dimensionless constant $a_{0}$ numerically as $a_{0}=3.5.$ Figure 4
demonstrates that two dipole-type solitons may exist in the fiber medium. As
seen, this double dipole-type soliton keeps its profile over a distance of
forty normalized lengths. In addition to double dipole-type soliton modes,
we have found that families of multiple-hump solutions can also be generated in
the system, including coupled three, four, and N dipole-type soliton pulses.

\section{Numerical stability analysis}

A distinguishing property of localized pulses is their stability to
perturbations, as only stable solitons can be observed experimentally and
utilized in physical applications. It is therefore important to analyze the
stability of the obtained multiple-hump solutions with respect to small
perturbations. Here, we take the numerically found double-hump quartic and
dipole soliton solutions as examples to perform numerical experiments of the
model (\ref{1}). The numerical evolution of double-hump quartic and dipole
soliton solutions under the perturbation of 10\% white noise are depicted in
Figs. 5 and 6, respectively. These results show that the multiple-hump
soliton modes can propagate stably in the fiber system under the initial
perturbation of the additive white noise. Hence we can conclude that the
novel multi-soliton modes we obtained are stable.

\section{Conclusions}

We have presented the first analytical and numerical demonstration of the
existence of multi-humped soliton pulses in a highly dispersive optical
fiber exhibiting a Kerr nonlinearity. We revealed that in such nonlinear
medium, the presence of cubic and quartic dispersions may lead to the
coupling of quartic or dipole solitons into localized multi-hump pulses. The
dynamics of the newly found solutions in the fiber material have been found
to be very well modeled by the extended nonlinear Schr\"{o}dinger equation
incorporating the contributions of second-, third-, and fourth-order
dispersions and self-phase modulation. The results demonstrate surprisingly
that both quartic and dipole types of double-, triple-, and multi-humped
soliton solutions can be formed in the fiber system. We have also
demonstrated numerically that such soliton modes are stable with respect to
small perturbations, thus implying that they can be utilized for
transmission. To our knowledge, the multi-hump optical solitons obtained
here for the extended NLSE are firstly reported in this work.

The discovery of such coupled multi-quartic and multi-dipole solitons
represent an important advance in nonlinear optics. It should be mentioned
that the newly found localized multi-solutions could find important
application not only in optical fibers but also in other physical systems
for which the underlying equation is applied for describing the wave
dynamics. In future research problems, we shall take into consideration the
absorption or amplification effects to expand the applicability of obtained
multi-soliton solutions.

\end{document}